\documentclass[11pt]{article}
\pdfoutput=1

\usepackage{amsmath,amssymb,bm,color}
\usepackage{appendix}
\usepackage{graphicx}
\usepackage{xcolor}
\usepackage{cancel}
\usepackage{jcappub}
\usepackage[T1]{fontenc} 
\usepackage{slashed}
\usepackage{soul}

\newcommand{\be}{\begin{equation}}
\newcommand{\ee}{\end{equation}}
\newcommand{\bea}{\begin{eqnarray}}
\newcommand{\eea}{\end{eqnarray}}

\title{\boldmath Recycled Dark Matter}

\author[a]{Thomas C. Gehrman,}

\author[b]{Barmak Shams Es Haghi,}

\author[a]{Kuver Sinha,}

\author[a]{and Tao Xu}
\affiliation[a]{Homer L. Dodge Department of Physics and Astronomy, University of Oklahoma, Norman, OK 73019, USA}
\affiliation[b]{Texas Center for Cosmology and Astroparticle Physics, Weinberg Institute for Theoretical Physics, Department of Physics, The University of Texas at Austin, Austin, TX 78712, USA}

\emailAdd{thomas.gehrman@ou.edu}

\emailAdd{shams@austin.utexas.edu}

\emailAdd{kuver.sinha@ou.edu}

\emailAdd{tao.xu@ou.edu}

\abstract{We outline a new production mechanism for dark matter that we dub ``recycling'': dark sector particles are kinematically trapped in the false vacuum during a dark phase transition; the false pockets collapse into primordial black holes (PBHs), which ultimately evaporate before Big Bang Nucleosynthesis (BBN) to reproduce the dark sector particles. The requirement that all PBHs evaporate prior to BBN necessitates high scale phase transitions and hence high scale masses for the dark sector particles in the true vacuum. Our mechanism is therefore particularly suited for the production of ultra heavy dark matter (UHDM) with masses above $\sim 10^{12}\,{\rm GeV}$. The correct relic density of UHDM is obtained because of the exponential suppression of the false pocket number density. Recycled UHDM has several novel features: the dark sector today consists of multiple decoupled species that were once in thermal equilibrium and the PBH formation stage has extended mass functions whose shape can be controlled by IR operators coupling the dark and visible sectors.} 

\begin{document}
\hfill{\small UTWI-35-2023}

\maketitle
\flushbottom

\section{Introduction}
\label{sec:intro}

Thermal freeze-out has for a long time been the preferred history of the cosmological evolution of dark matter (DM). Its most outstanding feature is the fact that DM with weak scale interactions and a thermal history reproduces the observed relic abundance. Given that new physics is expected to appear at the weak scale for completely unrelated reasons (for example the hierarchy problem), this fact presented itself as a ``miracle'' for several decades, leading to DM searches over numerous facilities geared at the weak scale thermal candidate.

The failure of said candidate to materialize has prompted a flurry of DM and early Universe model-building in recent years. A part of the community has focused on  changing the equation of state of the early Universe (for example, by introducing an early period of  matter domination \cite{Allahverdi:2020bys, Kane:2015jia}), thereby altering the thermal evolution of DM and opening up vast regions of hitherto disallowed parameter space. Another part of the community has focused on newer mechanisms by which DM achieves the observed relic density, some of which are necessitated by the vastly different DM mass and couplings compared to  the weak scale \cite{Cooley:2022ufh, Baryakhtar:2022hbu}. Model building to go beyond the WIMP paradigm for achieving the correct relic abundance has been studied by many authors; for a flavor of recent studies, we refer to  \cite{DEramo:2010keq, Hochberg:2014dra, DAgnolo:2015ujb, Dror:2016rxc, Kuflik:2015isi, Cline:2017tka, DAgnolo:2017dbv, Garny:2017rxs, Smirnov:2020zwf, Fitzpatrick:2020vba, Frumkin:2021zng}.

The purpose of this paper is to explore one such new mechanism for the production of DM, which we dub ``recycled dark matter''. Recycling works as follows. A single or multiple (say, two) species of particles is/are trapped in a false vacuum region during a first order phase transition (FOPT) due to kinematics (they are very heavy in the true vacuum, and practically massless in the false vacuum).  The species, which one can imagine to constitute a dark sector, maintains thermal equilibrium in its own sector and  kinetic equilibrium with the Standard Model (SM) before the FOPT. The rate at which the trapped species of the dark sector leaks into the true vacuum or annihilates is suppressed enough that at  some point, the false vacuum collapses into a primordial black hole (PBH) (this condition is what requires the introduction of multiple species in the dark sector, as we shall see). For phase transition temperatures high enough, the mass of the PBHs thus formed is small and they all evaporate before Big Bang Nucleosynthesis (BBN). When they evaporate, they produce (``recycle'') the dark sector that originally went into forming them. At this stage, however, the dark sector species are thermally decoupled, ultra-heavy (with their true vacuum mass), and constitute a cold dark sector consisting of multiple DM species.  

One can ask: beyond the novelty of the recycling mechanism, does this production mechanism solve any outstanding issues for DM models? Our answer is in the affirmative: it presents a very useful way to think about the production of DM heavier than the Griest-Kamionkowski bound~\cite{Griest:1989wd} -- the so-called ultra heavy DM (UHDM) candidates. The logic is as follows. Ultimately, recycling arises from the fact that the PBHs all Hawking evaporate before BBN, which necessitates light PBHs and therefore phase transitions at a high temperature. But the temperature during the phase transition is set by the (also therefore high) vacuum expectation value of whatever scalar field is responsible for the symmetry breaking, which in turn sets the mass-scale of the dark sector species. Thus, the recycling mechanism is most naturally suited for UHDM candidates. 

One can then ask: how is the correct relic density obtained in this scenario? Typically, UHDM candidates beyond the Griest-Kamionkowski bound require an exponential suppression of their number density to achieve the correct relic density. This can be achieved by various mechanisms of which we mention a few (for a recent review, we refer to the Snowmass report \cite{Carney:2022gse} and references therein): entropy dilution \cite{Berlin:2016gtr}, an enhancement of the annihilation cross section (for example through Sommerfeld enhancement \cite{Hisano:2006nn}), specific multi-species annihilation patterns \cite{Kim:2019udq, Berlin:2017ife, Kramer:2020sbb, Frumkin:2022ror}, the suppression of the number density through gravitational production (for example the case of WIMPzillas \cite{Chung:1998ua}), various avatars of freeze-in and  Boltzmann suppression (for example being filtered \cite{Chway:2019kft, Baker:2019ndr} or squeezed out \cite{Asadi:2021yml} during phase transitions), and Hawking radiation of PBHs~\cite{Lennon:2017tqq, Morrison:2018xla, Hooper:2019gtx}. In contrast, for recycled DM the exponential suppression of the DM number density arises from the distribution of false vacuum pockets shown in Eq.~\eqref{eq:dnpocketdRstar}. 

It is important to contrast recycling with other UHDM production mechanisms that rely on phase transitions. The mechanisms that are most closely related are filtered DM and dark quark nuggets and Fermi-balls \cite{Bai:2018dxf,Hong:2020est,Kawana:2022lba}\footnote{Mechanisms that rely on super-cooling \cite{Azatov:2021ifm} or bubble wall collision (babyZillas \cite{Falkowski:2012fb}) are quite different.}. In filtering, the DM candidates escape from the false vacuum to the true vacuum where their number density is Boltzmann suppressed while the remaining DM particles in the false vacuum annihilate away. In recycling, on the other hand, the DM candidates cannot filter out of the false vacuum as efficiently, since their annihilation products are also massive in the true vacuum.  Trapped in the false vacuum, the DM particles annihilate into each other and the annihilation products are also too heavy to escape efficiently. The dark sector maintains its own thermal equilibrium  while evolving separately to a higher temperature than the SM sector. Finally, a PBH forms when the trapped dark sector is compressed to below its Schwarzschild radius.

\begin{figure}[h]
  \centering
    \includegraphics[width=0.95\textwidth]{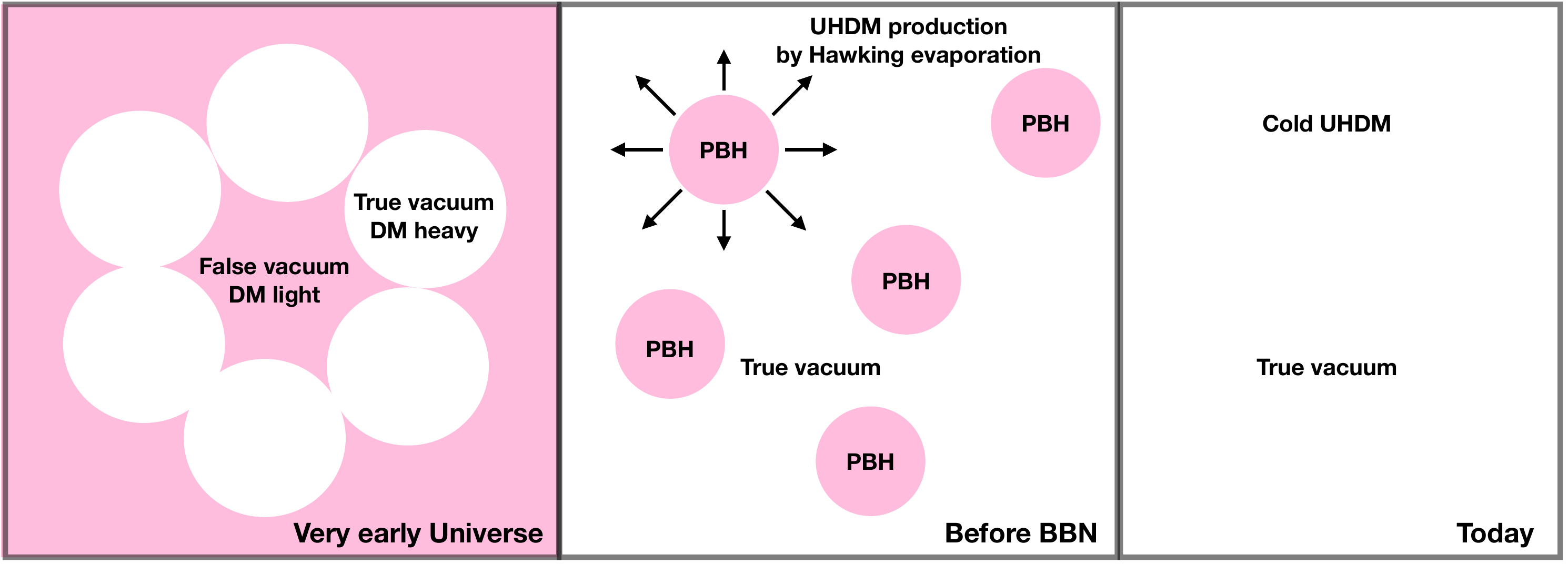}
  \caption{ 
  {The general scheme of recycled DM. \textbf{Left:} In very early Universe, a dark scalar field develops a vacuum expectation value (vev) and gives rise to a FOPT which is succeeded by formation of true vacuum bubbles. Dark sector particles, which are in kinetic equilibrium with the SM particles, are  assumed to be massless in the false vacuum while they acquire a very large mass in the true vacuum. Therefore, they remain trapped in the false vacuum. \textbf{Middle:} False vacuum pockets (containing dark sector particles) that are large enough collapse into light PBHs and eventually evaporate before BBN to emit all the particles in the spectrum including dark sector particles. \textbf{Right:} The dark sector particles emitted by PBHs get diluted by the expansion of the Universe until today to explain the observed abundance of DM.}
  }
  \label{fig:scheme}
\end{figure}

The distinction of recycling with dark quark nuggets or Fermi-balls is that in recycling, the PBHs all completely evaporate and reproduce the dark sector that originally collapsed (i.e., the UHDM are not compact objects but simply regular GUT-scale particles formed from an evanescent intermediate PBH stage). The main stages are illustrated in Fig.~\ref{fig:scheme}.

Our main goal in this paper is to offer a proof of concept of recycling, using a simple DM sector consisting of a real scalar $\phi$ and a Dirac fermion $\chi$ as a template. The scalar undergoes spontaneous symmetry breaking and acquires a vev $\sim 10^{11} - 10^{15}$ GeV; consequently both the scalar and the fermion acquire large masses at this scale in the true vacuum. We compute the filtering rate and position ourselves in a region of parameter space where the filtered DM contribution to the relic density is small. We go on to study the conditions for PBH formation in false vacuum pockets and the ensuing yield of the scalar $\phi$ and fermion $\chi$ from Hawking evaporation, obtaining the parameter space where the correct relic density will be produced.

There are two novel features of recycled DM that we want to emphasize. The first pertains to the PBH side of the story, while the second pertains to the UHDM side. On the PBH side, recycling features \textit{extended PBH mass functions whose shape is controlled by IR physics\footnote{We denote by ``IR physics'' any physics that does not directly depend on the inflationary era, which is where PBH mass functions have typically been modelled. In our scenario, the PBH mass function can be tuned in ordinary effective field theory, without appealing to Planck-scale physics.}}. The reason is as follows. Since the products of DM annihilation in the false vacuum remain trapped there, pockets of all sizes ultimately collapse to PBHs (this can be contrasted to the situation in \cite{Chway:2019kft, Baker:2019ndr}). The lower cutoff in the PBH mass function is provided by the strength of the interaction that maintains the dark and SM sectors at the same temperature at the onset of the FOPT and facilitates dark sector annihilation into SM particles that escape the pocket. The strength of this interaction provides parametric control over the shape of the mass function, which we display in Fig.~\ref{fig:LambdaTstar}. This furnishes an example of the shape of PBH mass functions being controlled by the cutoff scale and coupling of effective operators in the IR. 

On the UHDM side, the novel feature is the following: recycling requires multiple dark sector species in thermal equilibrium prior to PBH formation, which are then thermally decoupled from each other after Hawking evaporation. Depending on the mass of the PBHs which produce them and also the interactions with the SM, they may or may not come back to kinetic equilibrium with the SM bath, but they cannot reach chemical equilibrium. Some (at least one) of these species will  have long enough lifetimes to persist to the current Universe, and therefore the cold DM comprising the relic density today is typically composed of multiple  UHDM candidates.  This provides an  example of a multi-component dark sector whose components were  once in thermal equilibrium but are now thermally decoupled.

Our paper is organized as follows. In Section \ref{filtereddm}, we introduce our template model and perform a calculation of the trapping fraction. We also discuss why recycling models require multiple particle species in the dark sector. In Section \ref{thermalhistory}, we discuss the thermal history of the dark sector, the PBH formation process, and the PBH mass function. In Section \ref{uhdmfrompbh}, we discuss the production of UHDM from PBH evaporation and show the regions of the parameter space where recycling gives the observed relic density. We also comment on the annihilation of DM after production from PBH evaporation and the contribution of filtered DM to the final relic density. We end with our conclusions.

\section{Trapped Dark Sector during FOPT} \label{filtereddm}

In this Section, we introduce the template dark sector Lagrangian that we will be working with. We then discuss the novel trapping behavior of the dark sector during a FOPT. This enables us to obtain the mass scales where the recycling effect is dominant for generating the observed DM abundance. In addition, we compare dark sector models that host black hole formation from the over-density of trapped particles, and introduce the idea of multi-particle dark sector.    

\subsection{Dark Sector Model}
\label{sec:TrappedDarkSector}
To provide an illustrative example of recycling, we introduce a scalar $\phi$ and a Dirac fermion $\chi$ with interactions as follows:
\be
\mathcal{L} \supset - y_\chi \phi  \bar{\chi} \chi + \mu^2 \phi^2 - \lambda \phi^4 + \mathcal{L}_{\rm SM-DS}
\label{eq:Lagrangian}
\ee
Here we will focus our discussion on the dark sector while refraining from making any assumptions about the specifics of the dark sector-SM interactions $\mathcal{L}_{\rm SM-DS}$, which we only assume is strong enough to keep the two sectors with the same temperature before the FOPT. In the rest of this subsection, we use the subscript $d$ to indicate a particle belongs to the dark sector, without specifying its detailed particle nature other than those affecting the trapping process. The result applies to both $\chi$ and $\phi$ particles in Eq.~\eqref{eq:Lagrangian}. We will assume that $\phi$ undergoes a FOPT and choose a selection of the parameters $\{\mu, \lambda, y_\chi\}$ such that $m_\chi = m_\phi = y_\chi \langle \phi \rangle$ in the true vacuum. One could also endow the dark fermionic sector with an additional flavor structure: $y_{ij}$. In the rest of the paper, we will consider a single flavor, reserving the multi-flavor case for future work.

Now we calculate the trapping rate of dark sector particles during the FOPT from the time $T=T_\star$ when false vacuum regions nucleate and keep shrinking as bubble wall are moving at a velocity of $v_w$. It is the gap between particle mass values on the true vacuum side and the false vacuum side of the bubble wall that determines the trapping rate. Unless sitting on the exponentially suppressed Boltzmann tail, a DM particle with a smaller mass in the false vacuum cannot enter the true vacuum due to energy momentum conservation. We keep to notation for bubble wall velocity in the $+z$ direction. To obtain the probability of particles being bounced back from the bubble wall, we can first calculate the population $\Delta N_{\rm in}$ of particles that satisfy the pass-through condition $\mathcal{T}(\vec{p})$ in front of the bubble wall area $\Delta A$ within time duration $\Delta t$ \cite{Chway:2019kft},
\bea
\frac{\Delta N_{\rm in}}{\Delta A} 
&=& \frac{g_d}{(2 \pi)^3} \, \int d^3\vec{p} \, \int^{r_0-\frac{p_z \, \Delta t}{|\vec{p}|}}_{r_o} dr \, \mathcal{T}(\vec{p}) \, \Theta(-p_z) \, f(\vec{p};\vec{x}) \nonumber\\
&=& \frac{g_d}{(2\pi)^3}\int d^3\vec{p} \, \Theta(-p_z-m_d) \, \Theta(-p_z) \, (-\frac{p_z \, \Delta t}{|\vec{p}|}) \, f(\vec{p};\vec{x}).
\eea
$\Theta(.)$ is the Heaviside step function. A particle can penetrate the wall if its momentum $\vec{p}$ projected in the perpendicular direction towards the bubble wall $-p_z$ is larger than the mass gap, meaning $\mathcal{T}(\vec{p})=\Theta(-p_z-m_d)$.
For the dark sector particle, $g_d$ is its degree of freedom, $m_d$ is its mass in the true vacuum side of the bubble wall, and $f(\vec{p};\vec{x})$ is its phase space distribution in the rest frame of the bubble wall. The absolute value of the penetrated particle flux can be defined in the wall rest frame as $J_w\equiv\frac{dN}{dAdt}\simeq\frac{\Delta N}{\Delta A \, \Delta t}$, which can be written as, 
\bea
J_w 
&=& \frac{g_d}{(2\pi)^3} \int d^3\vec{p} \, \Theta(-p_z-m_\chi) \, \Theta(-p_z) \, f(\vec{p};\vec{x}) \, (-\frac{p_r}{|\vec{p}|}) \nonumber\\
&=& \frac{g_d}{(2\pi)^3} \int^{\pi}_{0} d\theta \, \sin\theta \, \int^{2\pi}_0 d\phi \, \int^{\infty}_{0} dp \, p^2 (-\cos\theta) \, \Theta(-p\cos\theta-m_d) \, \Theta(-\cos\theta) \, f(\vec{p}).\nonumber \\
\label{eq:Jwall}
\eea
In the second line, we simplify the particle momentum integral in spherical coordinate system. We also assume the phase space distribution is homogeneous such that $f(\vec{p};\vec{x})=f(\vec{p})$ only depends on the particle momentum. 

The phase space distribution $f(\vec{p})$ can be obtained by imposing a Lorentz transformation on the particle equilibrium distribution from the local plasma rest frame to the bubble wall rest frame. The particle would be boosted if there was a relative motion between the bubble wall and the plasma in front of the wall. We assume the temperature of dark sector particles is $T$. The bulk of local plasma is moving with a velocity $\vec{\tilde{v}}$ in the $-\hat{z}$ direction towards the wall. Here bulk refers to the overall motion of the particles. Then the phase space distribution in the wall rest frame is
\bea
f(\vec{p})
&=& \frac{1}{e^{\tilde{\gamma} \, (E \, - \, \vec{\tilde{v}} \cdot \vec{p})/T} \pm 1 } \nonumber\\
&\simeq& \frac{1}{e^{\tilde{\gamma} \, (p \, + \, \tilde{v} \, p \, \cos\theta )/T}}.
\label{eq:PhaseSpaceDistribution}
\eea
The Lorentz factor from $\tilde{v}$ is $\tilde{\gamma}=1/\sqrt{1-\tilde{v}^2}$. In the last line, we approximated the phase space distribution with a Boltzmann distribution of massless particles. The difference coming from this approximation for a Bose-Einstein distribution and a Fermi-Dirac distribution is found to be $\mathcal{O}(0.2)$ in \cite{Chway:2019kft}. Note that particles are massless in the false vacuum side of the bubble wall. Plugging Eq.~\eqref{eq:PhaseSpaceDistribution} into Eq.~\eqref{eq:Jwall}, we obtain:
\bea
J_w
&=& \frac{g_d}{(2\pi)^2} \int^{-1}_0 d\cos\theta \, \cos\theta \int^{\infty}_{-\frac{m_d}{\cos\theta}} dp \, \frac{p^2}{e^{\tilde{\gamma} \, (1 \, + \, \tilde{v} \, \cos\theta ) \, p / T}} \nonumber\\
&=& \frac{g_d \, T^3 \, \left(1 \, + \, \tilde{\gamma} \, m_d \, (1-\tilde{v})/T \right)}{4 \, \pi^2 \, \tilde{\gamma}^3 \, (1-\tilde{v})^2} \, e^{-\tilde{\gamma} \, m_d \, (1-\tilde{v})/T}.
\label{eq:JwallFinal}
\eea
Eq.~\eqref{eq:JwallFinal} agrees with previous studies \cite{Chway:2019kft} up to the sign difference to account for the direction of the flux. One can see the penetrated flux is exponentially suppressed by the ratio of particle mass in the true vacuum and particle temperature in the false vacuum, indicating that we can trap most particles inside false vacuum regions.

In the end, the number density of particles penetrated into the true vacuum region seen in the global plasma frame is
\bea
n_{d}^{{\rm filtered}} = \frac{J_w}{\gamma_w \, v_w}.
\label{eq:ndfiltered}
\eea
Here the bubble wall velocity in the global plasma frame is $v_w$, and $\gamma_w=1/\sqrt{1-v_w^2}$.
One should note the difference between velocities $v_w$ and $\tilde{v}$, and between the Lorentz factors $\gamma_w$ and $\tilde{\gamma}$. Quantities represented with tilde are defined in the bubble wall rest frame, and quantities represented without tilde are defined in the global plasma frame. The Lorentz factor $\gamma_w$ in the denominator accounts for the time dilation effect experienced by the observer in the global plasma frame. The $n_{d, {\rm filtered}}$ represents the number density of the dark sector species that can enter the true vacuum regardless of the sudden mass increase at the bubble wall. On the other hand, the fraction of particles that are trapped is defined
\bea
F^{\rm trap}=1-{n_{d}^{\rm filtered}}/{n_d},
\eea
where $n_d$ is the equilibrium number density of the dark sector particle.

\begin{figure}[h]
  \centering
    \includegraphics[width=0.45\textwidth]{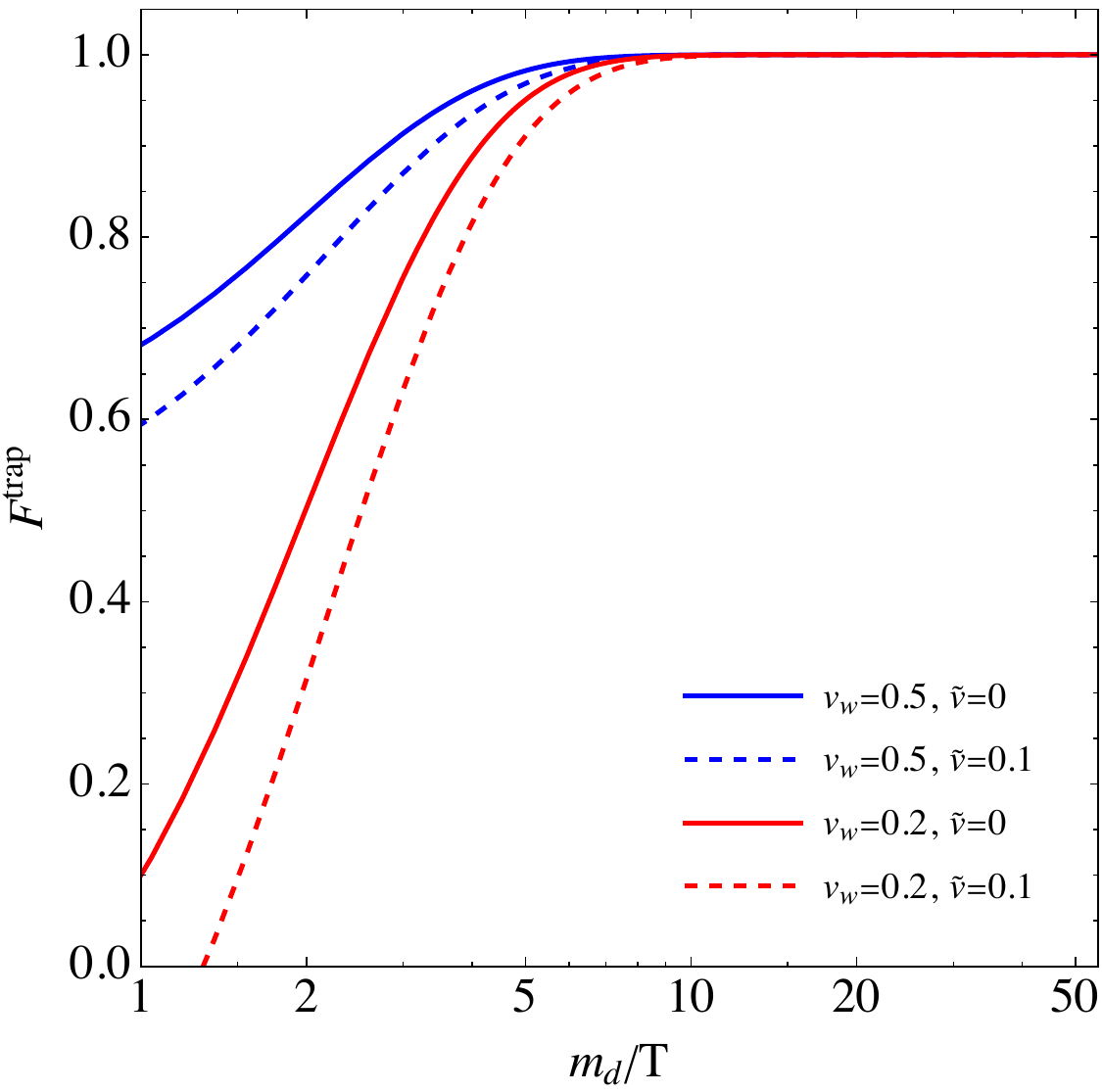} 
  \caption{The trapping fraction $F^{\rm trap}$ as a function of $m_{d}/T$ for different relative velocity of particle bulk motion in front of the bubble wall $\tilde{v}=0$ (solid) and $\tilde{v}=0.1$ (dashed). $m_d$ denotes either $m_\chi$ or $m_\phi$. The bubble wall velocity in the global plasma frame is shown in different colors $v_w=0.5$ (blue) and $v_w=0.2$ (red).}
  \label{fig:TrappingFrac}
\end{figure}

In Fig.~\ref{fig:TrappingFrac}, we show the fraction of trapping $F^{\rm trap}$ during FOPT as a function of the particle mass-temperature ratio $m_d/T$. We show the limit $\tilde{v}=0$ in solid curves where the maximum trapping rate is achieved with the assumption that the bulk gains an overall velocity to move relatively at rest with respect to the bubble wall. As a comparison, we show the case of $\tilde{v}=0.1$ in dashed curves where the trapping rate is suppressed by the Lorentz boost on the particle phase space distribution in Eq.~\eqref{eq:PhaseSpaceDistribution}. We also show the bubble wall velocities $v_w=0.5$ (blue) and $v_w=0.2$ (red). The trapping rate is higher with a larger $v_w$ as a result of the time dilation effect in Eq.~\eqref{eq:ndfiltered}.

\subsection{Why a Multi-particle Dark Sector is Required}

Most models of PBH formation from  phase transitions in the recent literature invoke a single fermion species that gains mass from symmetry breaking, while the scalar responsible for the symmetry breaking remains massless. In such scenarios, the dark sector ultimately consists of only one species (the fermion). In this subsection, we discuss the challenges of implementing the recycling mechanism for such cases.

The main challenge  lies in the annihilation of the fermion species $\chi$ to the scalar $\phi$ via $\bar{\chi}\chi \rightarrow \phi$  and $\bar{\chi}\chi \rightarrow \phi \phi$ allowed by the Yukawa coupling $y_\chi \phi \bar{\chi} \chi$. For massless scalars, the $\phi$ particles thus produced escape into the true vacuum. This effectively leaks out the energy density of the false vacuum pockets and inhibits PBH formation. One may be able to suppress this annihilation by reducing the Yukawa coupling, but that introduces its own problems as far as recycling is concerned. Indeed, the scaling arguments for the Yukawa coupling forwarded in \cite{Baker:2021sno} are instructive in this regard. The annihilation $\bar{\chi}\chi \rightarrow \phi$ scales as $y^2_\chi T_*$ while the Hubble length scales as $T^{-2}_*$ in a radiation-dominated Universe. Therefore, to keep the number of annihilations per Hubble length invariant, the Yukawa should scale as $y_\chi \, \sim \, 10^{-5}\sqrt{T_*/{\rm PeV}}$. This gives a very suppressed Yukawa at the temperatures we are interested in. As we will see in Fig.~\ref{fig:betaFOPTmOverT}, recycling requires 
\be
\langle \phi \rangle \sim \mathcal{O}(55-60) \times T_* \times (1/y_\chi)\,\,.
\label{eq:filterDM}
\ee
A small Yukawa coupling therefore would imply a huge hierarchy between $\langle \phi \rangle$ and $T_*$.

For phase transitions at higher temperatures,  the $\bar{\chi}\chi \rightarrow \phi \phi$ annihilation process becomes operational, and the appropriate scaling is $y^4_\chi T_*$. The benchmarks studied in \cite{Baker:2021sno} where successful PBH formation could occur included $y_\chi = 0.3$ for $T_* = 10^{15}$ GeV. From Eq.~\eqref{eq:filterDM}, it  is clear that this would require a large hierarchy between $\langle \phi \rangle$ and $T_*$. We therefore see that recycling requires large hierarchies between $\langle \phi \rangle$ and $ T_*$, if the scalar $\phi$ is allowed to escape into the true vacuum. One option to ameliorate the large hierarchy could be to increase $y_\chi \sim \mathcal{O}(\sqrt{4\pi})$. However, this would only increase the annihilation of $\bar{\chi}\chi$ into $\phi\phi$. Therefore, a heavy $\phi$ particle that is itself trapped in the false vacuum is necessary.

A different strategy could be pursued if one insisted on keeping $\phi$ massless. This would be to allow the symmetric component of $\chi$ to annihilate away to $\phi$ which escapes, but to have the asymmetric component remain in the false vacuum. This is the option followed in scenarios of Fermi-ball formation, about which we now make a few comments.
The asymmetry in the distribution of the $\chi$ particles can be generated with a chemical potential as in usual asymmetric DM models \cite{Kaplan:2009ag, Zurek:2013wia}, or accidentally from the Gaussian statistics of the $\chi$ relic number density \cite{Asadi:2021yml,Asadi:2021pwo}. 
It should be noted that both the chemical potential generated asymmetric abundance and the accidentally generated asymmetric abundance are usually discussed in the context of particles that are already out of equilibrium.
In our opinion, the application of these asymmetry generation mechanisms to the massless fermionic species which is still in equilibrium in the false vacuum pocket needs further study. Such further investigation would  confirm the generation of a dominant $\chi$ population over the $\bar\chi$ particle in the false vacuum; currently, however, we do not pursue this route.

One could imagine that the presence of the $\phi$-mediated attractive force might enhance PBH formation before dark sector annihilation over-dilutes the energy density in the false vacuum. The bubble wall keeps compressing the $\chi$ and $\phi$ particles left within the pocket until the mean separation between $\chi$ particles becomes shorter than the range $m_{\phi}^{-1}$ of the strong Yukawa interaction mediated by $\phi$. Since the Yukawa force is attractive, the additional attraction could lead to the collapse of $\chi$ particles into a PBH before the fermion degeneracy pressure becomes relevant. Moreover, the PBH formation could happen for either symmetric or asymmetric distribution between $\chi$ and $\bar{\chi}$ particles. The asymmetric distribution case has been studied for the collapse of Fermi-balls due to an internal Yukawa force in \cite{Kawana:2021tde,Huang:2022him}. A detailed simulation of the collapse process triggered by the Yukawa interaction is necessary to determine the required strength of the Yukawa force as well as the collapse criteria (for example, how the Schwarzschild radius compares to the particle separation and the Yukawa force range). We do not comment on this possibility further in the present work.

Finally, there is one more reason why a multi-particle dark sector is required for recycling models. Typically, one requires the dark sector trapped in the false vacuum to maintain kinetic equilibrium with the SM up to the stage when the phase transition occurs. This involves couplings between the dark sector and the SM sector through higher dimensional operators suppressed by some cutoff scale. These operators can subsequently induce decays of the dark sector particles after they are produced from Hawking evaporation, jeopardizing their status as cold DM today. In the case of a single dark sector particle, this eventuality can impose severe constraints on the mass of the DM and the cutoff scale. Such a situation  is avoided for multiple particles in the dark sector; even if one species decays, others can constitute cold DM in the current Universe.

We point out an important caveat here. The dark sector is required to have multiple components during the PBH formation stage and also during Hawking evaporation for reasons we have just discussed. However, after evaporation, the species that is responsible for establishing kinematic equilibrium with the SM may typically mediate early decays of other species, since we are considering UHDM. In our template model, we will choose the decaying particle to be $\phi$, and the cold DM in the current Universe will be constituted by (the possibly many flavors of) the $\chi$ particles.

\section{Thermal History and PBH Formation}\label{thermalhistory}

In this section, we discuss the thermal history of the dark sector whose particles are trapped during the process described in Sec.~\ref{sec:TrappedDarkSector}. The trapped particles generate local over-dense regions, and if the trapping process lasts long enough, these regions collapse into PBHs. We examine the condition of forming PBHs and calculate the their mass function for our scenario\footnote{For general reviews of PBHs, we refer to \cite{Escriva:2022duf, Carr:2020xqk}.}. As we shall see, the term $\mathcal{L}_{\rm SM-DS}$ in Eq.~\eqref{eq:Lagrangian} that guides the interaction of the two sectors will play a critical role in this discussion.

\subsection{Thermal History of the Dark Sector}

We start with the term in the Lagrangian that mediates the interaction between the hidden sector and the SM:
\be
\mathcal{L}_{\rm SM-DS} = \frac{\alpha_{\Lambda}}{\Lambda} \bar{\chi}\chi H^\dagger H \,\,,
\label{eq:EFTLambda}
\ee
where $\alpha_\Lambda$ is the coupling constant of the portal interaction, $H$ is the SM Higgs and $\Lambda$ is a cut-off scale that we will take to be near the GUT scale. Other portal operators would work as well; we take Eq.~\eqref{eq:EFTLambda} to illustrate our main ideas. The thermal evolution of the dark sector is tracked from temperatures  $T\gg\Lambda$. The dark sector and the SM are in equilibrium at the onset of the cosmic expansion. Kinetic equilibrium can be maintained until $T\simeq\Lambda$ provided that the scattering rate is larger than the Hubble expansion rate, $\Gamma_{\rm sca}=n_{\rm SM} \, \alpha_\Lambda^2/\Lambda^2 \gtrsim H(\Lambda)$. Here $n_{\rm SM}$ is the number density of SM particles that a dark sector particle can scatter with. We can derive the lower limit on $\alpha_\Lambda$ to maintain kinetic equilibrium, as a function of the energy scale $\Lambda$,
\be
\alpha_\Lambda \gtrsim 0.17\times\,\left( \frac{g_{\star}+g_\phi+\frac{7}{8}g_\chi}{106.75+4.5} \right)^{{1}/{4}} \, \left(\frac{\Lambda}{10^{16}~{\rm GeV}}\right)^{{1}/{2}}\,\,.
\ee
Here $g_\star$ is the effective number of relativistic degrees of freedom for SM.
The same interaction also mediates the annihilation between SM and dark sector particles $\bar{\chi}\chi\to {\rm SM}$, $\phi\phi \to {\rm SM}$. When $T \gtrsim \Lambda$, the annihilation cross section scales as $\sigma_{\rm ann}= \alpha_\Lambda^2/T^2$, thus the annihilation rate $\Gamma_{\rm ann}=n_{\chi} \, \alpha_\Lambda^2/T^2\propto T$ is proportional to the temperature of the dark sector.  Below the scale $\Lambda$, the cross section for the annihilation is assumed to be constant: $\sigma_{\rm ann}=\alpha_\Lambda^2/\Lambda^2$. Therefore the annihilation rate $\Gamma_{\rm ann}=n_{\chi} \, \alpha_\Lambda^2/\Lambda^2\propto T^3$ is proportional to $T^3$, which means the energy loss from the dark sector to the SM sector is more efficient at places where the dark sector temperature is higher.

When the temperature drops to $T\lesssim\Lambda$, the two sectors start to decouple from equilibrium, because the particle number density keeps dropping while the portal cross section saturates the constant value $\alpha_\Lambda^2/\Lambda^2$. This
suppresses the annihilation and scattering rate between the two sectors. Kinetic equilibrium may be kept longer than chemical equilibrium because of the larger  number of degrees of freedom for target particle species for the scattering process than $g_{\chi/\phi}$ for the annihilation process. In general, we expect the two sectors to evolve separately with their own temperatures after their decoupling. We assume that there is no reheating effect during $T_\star \lesssim T \lesssim \Lambda$, such that two sectors accidentally evolve to the same temperature when the FOPT starts. A model dependent analysis could determine the deviation from this assumption, but it will not change the conclusions of this study. Therefore, we take $T_{\chi}=T_{\phi}=T_{\rm SM}$ when we study the dynamics of the dark sector during the phase transition.

\subsection{Black Hole Formation}

Since the equilibrium between the dark sector and the SM sector can be achieved with the effective coupling in Eq.~\eqref{eq:EFTLambda}, we assume $T_\chi=T_\phi=T_\star$ where $T_\star$ is the nucleation temperature of the false vacuum. Inside the bubbles, the symmetry is broken and the masses of the dark sector particles are $m_\phi = m_\chi \sim y_\chi \langle \phi \rangle$. On the other hand, outside the bubble in the false vacuum pockets, the masses are their pole mass $m_\phi = 0$ and $m_\chi = 0$ with additional thermal corrections of $\mathcal{O}(T_{\star})$. Given that their mass gaps are much larger than their kinetic energy $m_{\chi/\phi} \gg T_\star$, dark sector particles are trapped inside the false vacuum and  heated up when they are bounced back from bubble walls. When the pocket radius shrinks to $R(t)$, the temperature evolves as $T_{\chi/\phi}\propto 1/R(t)$ like the behavior of radiation, while the particle number density increases as $n_{\chi/\phi}\propto (1/R(t))^3$. The enhanced energy density therefore causes the  Schwarzschild radius of the whole pocket mass to grow while the physical pocket radius decreases. When $R(t)$ coincides with the Schwarzschild radius of the pocket, the dense region will collapse into a PBH (the detailed calculation of the PBH mass function is in Sec.\ref{sec:PBHmassfunction}). The final pocket size at the time of collapse is found following the Schwarzschild criteria studied in \cite{Baker:2021nyl, Baker:2021sno} for our two component dark sector,
\bea
R(t_{\rm c})=\sqrt{\frac{\frac{7}{8}g_\chi+g_\phi}{g_\star}} \, \left( \frac{R_\star}{H_\star^{-1}} \right) \, R_\star.
\label{eq:Rtc}
\eea

\begin{figure}[h]
  \centering
\includegraphics[width=0.5\textwidth]{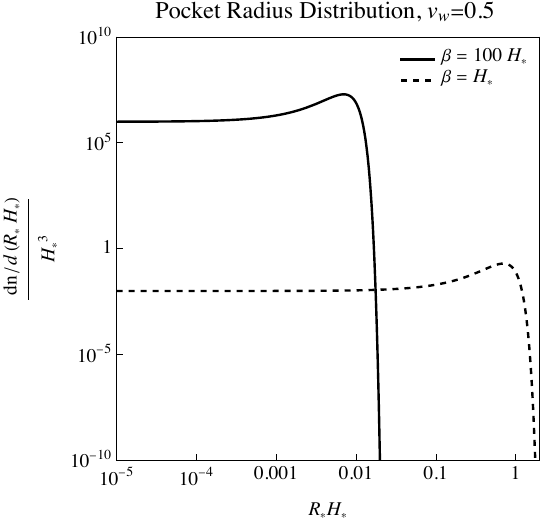}
  \caption{Pocket radius distribution for a fast FOPT, $\beta=100 H_\star$ (solid) and a slow one,  $\beta=H_\star$ (dashed) when $v_w=0.5$.}
  \label{fig:pocketdist}
\end{figure}

The pocket radius $R_\star$ at $T=T_\star$ is determined by the detailed phase transition model and could be verified with hydro-dynamical simulations of the nucleation of true vacuum bubbles. Since the initial condition of our calculation is set when most of regions are already in the true vacuum, the pocket shape is mostly spherical as a result of the surface tension of bubble walls. The initial pocket radius distribution has been studied for PBH formation in \cite{Lu:2022paj} with the requirement that a viable pocket should successfully shrink to form a PBH before the nucleation of new bubble seeds inside it. The pocket radius distribution is
\bea
\frac{dn_{\rm pocket}}{dR_{\star}} \simeq \frac{I_{\star}^4 \, \beta^4}{192 \, v_w^3}\, e^{4\beta R_{\star} / v_w-I_{\star} \, e^{\beta R_{\star}/v_w}} \left( 1 - e^{-I_{\star}e^{\beta R_{\star}/v_w}} \right).
\label{eq:dnpocketdRstar}
\eea
The pocket distribution is determined by the FOPT parameters. Here $v_{w}$ is the bubble wall velocity, $\beta$ is the inverse time scale of the FOPT process, and $I_\star = - \ln 0.29\simeq 1.238$ indicating $T_\star$ is the percolation temperature of false vacuum regions.
Fig.~\ref{fig:pocketdist} shows the pocket radius distribution for a fast FOPT corresponds to  $\beta=100 H_\star$ (solid), and a slow one, corresponds to  $\beta=H_\star$ (dashed) when $v_w=0.5$. For fast FOPTs, the pocket population, which later determines the PBH mass distribution, shifts to the smaller radii region as a result of the higher bubble nucleation rates that can split a large pocket into several smaller pockets.

The evolution of pockets of different sizes follows the same rule determined by the interaction of particles with the bubble wall. Since both the $\chi$ and $\phi$ particles gain a huge mass $m_{\chi/\phi} \gg T_\star$, the particles will not penetrate the bubble wall, but instead be reflected back from the bubble wall due to energy-momentum conservation. All pockets will be heated up following $T_{\chi/\phi}\propto 1/R(t)$ at the beginning, and if this process continues, all pockets are expected to collapse into PBHs spanning a extended mass range. 

If PBHs are able to form for all $R_\star$ values, then the total energy budget of the dark sector collapses into PBHs and therefore the initial energy density of PBHs would be of the order of a few percent of the energy density of the radiation (the ratio of the degrees of freedom in the dark sector to the degrees of freedom of the SM). Therefore, the total PBH energy density will dominate the energy density of the universe soon after their formation, resulting in an early matter-dominated era. Such an era will continue until  Hawking evaporation restores the cosmic evolution back to a radiation dominated era. However, energy loss from shrinking pockets to true vacuum regions can halt the evolution of pockets from becoming dense enough for PBH formation. Note that in Eq.~\eqref{eq:Rtc}, a smaller pocket has to shrink by a larger fraction of its initial radius to reach the Schwarzschild radius of the total enclosed energy. Since the final dark sector temperature is proportional to the factor $R_\star/R(t_{\rm c})\propto R_\star^{-1}$ according to Eq.~\eqref{eq:Rtc}, smaller pockets are heated to a higher temperature before their collapse. This means the energy loss will be more effective for pockets with smaller values of $R_\star$.  Such leakages can be achieved with the annihilation into particles that penetrate freely through the bubble wall. It was found in \cite{Baker:2021sno} that annihilation $\chi\bar{\chi}\to\phi, \phi\phi$ naturally stops PBH formation from pockets with $R_\star \lesssim 1.5\times H_{\star}^{-1}$ for their benchmarks,  assuming that  $\phi$ is not trapped inside the false vacuum. Since both dark sector particles are DM candidates in our study and are trapped in the false vacuum, annihilation within the dark sector will not change PBH formation and the threshold of \cite{Baker:2021sno} is not applicable. 

An alternative leakage channel is the annihilation through the operator in Eq.~\eqref{eq:EFTLambda}. To illustrate the idea, we assume that the annihilation process is $\chi + \bar{\chi} \to {\rm SM} + {\rm SM}$. The idea should apply to other portal models as well. The annihilate rate can be written as
\bea
\Gamma_{{\rm ann}} = n_{\chi} \, \frac{\alpha_\Lambda^2}{\Lambda^2},
\eea 
where the number density of $\chi$ could be tracked using
\bea
n_{\chi} &=& \frac{3}{4}\frac{\zeta(3)}{\pi^2} \, g_{\chi} T_\chi^3 \nonumber \\
&=& \frac{3}{4}\frac{\zeta(3)}{\pi^2} \, g_{\chi} \left( \frac{R_{\star}}{R(t)} \right)^3 \, T_\star^3.
\eea
We can write down the expected number of annihilation events throughout the successful collapse, which takes time $t_c\simeq R_\star/v_w$. Then the PBH formation condition is set by the range of $R_\star$ value where the expected number of annihilation events is smaller than $1$,
\bea
\Gamma_{\rm ann} \times t_c \lesssim 1.
\eea

This condition translates to a lower bound on $R_\star$ below which no PBH would form because of the active annihilation,
\bea
R_\star > 0.23 \, H_\star^{-1} \, \left(\frac{\alpha_\Lambda}{0.1} 
  \right) \, \left( \frac{0.5}{v_w} 
  \right)^{1/2} \, \left( \frac{g_\star+g_\phi+\frac{7}{8}g_\chi}{106.75+4.5} 
  \right)^{1/2} \, \left( \frac{T_\star}{\Lambda}
  \right) \, \left( \frac{M_{\rm Pl}}{T_\star} 
  \right)^{1/2}.
  \label{eq:RstarMin}
\eea

\begin{figure}[h]
  \centering
\includegraphics[width=0.5\textwidth]{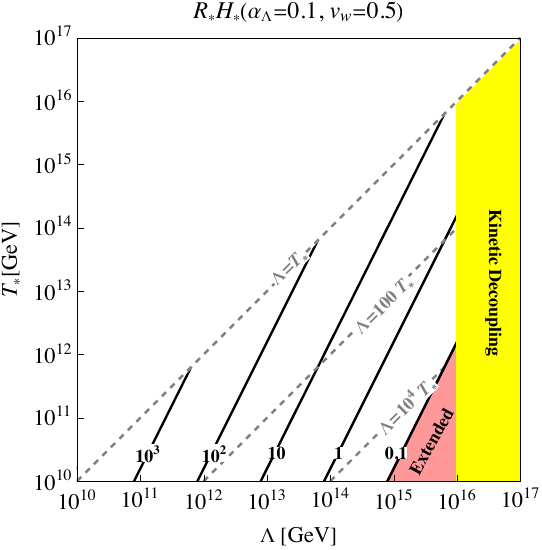}
  \caption{Lower bounds on $R_\star H_\star$ for $\alpha_\Lambda=0.1$ and $v_w=0.5$. For $\Lambda>10^{16}\,{\rm GeV}$ (yellow region) the dark sector drops out of kinetic equilibrium with the visible sector prior to the phase transition. Black solid lines show different values of $R_\star H_\star$. Within the red region, $R_\star H_\star$ becomes so small that the PBHs follow an extended mass function rather than a monochromatic one. The gray dashed contours are marked by different choices of $\Lambda$.}
  \label{fig:LambdaTstar}
\end{figure}

The minimal $R_\star$ normalized to the Horizon radius is shown on the $(\Lambda-T_\star)$ plane in Fig.~\ref{fig:LambdaTstar}, for $\alpha_\Lambda=0.1$ and $v_w=0.5$. The gray dashed contours depicts different values of $\Lambda$. The black solid lines are contours of minimal $R_\star$. For $R_\star \lesssim 0.1 \, H^{-1}_\star$, the red shaded region, the extension of the PBH mass function cannot be ignored.
In the yellow shaded region, for $\alpha_\Lambda=0.1$, the dark sector drops out of kinetic equilibrium with the visible sector.  
The valid parameter space is below the diagonal $\Lambda=T_\star$ dashed line as we assume a cutoff scale above the FOPT temperature. 

Eq.~\eqref{eq:RstarMin} is valid when the annihilation cross section remains a constant value. This means the dark sector temperature should not be heated to above $\Lambda$. We check the condition by combining Eq.\eqref{eq:Rtc} and \eqref{eq:RstarMin},
\bea
T_{\chi/\phi}(t_c)\lesssim 0.2 \, \Lambda  \, \left(\frac{0.1}{\alpha_\Lambda} 
  \right) \, \left( \frac{v_w}{0.5} 
  \right)^{1/2}\, \left ( \frac{T_\star}{10^{15}~{\rm GeV}} \right)^{1/2},
\eea
which confirms that the calculation is valid for the range of $T_\star$ of our interest for PBH formation. 

\subsection{Black Hole Mass Function}
\label{sec:PBHmassfunction}

In this section, we discuss the mass distribution of PBHs formed during a FOPT following \cite{Baker:2021sno, Xie:2023cwi, Gehrman:2023esa}.
The conditions for PBH formation in our model will be very similar, with one major difference: since the leakage of $\phi$ particles into the true vacuum is suppressed in our case, the size range of false vacuum pockets that can potentially collapse to PBHs is determined by the annihilation into SM particles, as shown in Eq.~\eqref{eq:RstarMin}.

We start with the  energy density at time $t_{\star}$ when the scalar and fermionic dark matter are in thermal equilibrium at some temperature $T_{\star}$ when the FOPT happens. The equilibrium energy densities of the scalar $\phi$ and the fermion $\chi$ are
\bea
\rho_{\chi}^{\rm eq} = \frac{7\pi^{2}}{240}g_{\chi} T_{\star}^{4} ~, \quad
\rho_{\phi}^{\rm eq} = \frac{\pi^{2}}{30}g_{\phi} T_{\star}^{4}.
\eea
The energy density of the trapped dark sector increases since the particles are being compressed by bubble walls. Assume $R_{\star}$ is the initial pocket radius at $t_\star$, and $R(t)$ is the pocket radius at a later time $t$. The energy density is found to follow the scaling \cite{Baker:2021sno}
\bea
\rho_{\chi,\phi}(t) = \rho_{\chi,\phi}^{\rm eq}\left(\frac{R_{\star}}{R(t)}\right)^{4}.
\eea
We define the total energy inside the pocket, 
\bea
M_{\rm pocket} = (\rho_{\phi}^{\rm eq} +  \rho_{\chi}^{\rm eq}) \left(\frac{R_{\star}}{R(t)}\right)^{4}\frac{4 \pi}{3}R(t)^{3},
\eea
and the Schwarzschild radius corresponding to the total enclosed mass,
\bea
r_{\rm c} = 2 \, G M_{\rm pocket}.
\eea
Here $G=M_{\rm Pl}^{-2}$ is the gravitational constant.

PBH formation occurs when $R(t)$ shrinks to the Schwarzschild radius $r_{\rm c} = R(t)$. We assume the energy of the whole dark sector inside the pocket is collapsed into the PBH. The PBH mass formed form a pocket of initial radius $R_\star$ is
\bea 
M_{ \rm PBH} = \left(\frac{2 \pi^{3}}{90 \, G} (g_{\phi} + \frac{7}{8}g_{\chi} )\right)^{1/2} R_{\star}^{2} \, T_{\star}^{2}.
\label{eq:MPBH}
\eea
The distribution of the PBH mass is determined by the distribution of the false vacuum pocket radius in Eq.~\eqref{eq:dnpocketdRstar} following the mass-radius relation given in Eq.~\eqref{eq:MPBH}. The lower bound on the pocket size, i.e., Eq.\eqref{eq:RstarMin}, translates into a lower bound on the mass of PBHs. Therefore the PBH distribution, including this lower bound, can be written as:
\begin{eqnarray}
   \nonumber \frac{dn_{\rm PBH}}{dM_{\rm PBH}} &=&
    \frac{I_{\star}^4 \, \beta^4M_{\rm PBH}^{-1/2}}{384 \, T_{\star}\, v_w^3\left(\frac{2 \pi^{3}}{90 \, G} \left(g_{\phi} + \frac{7}{8}g_{\chi} \right)\right)^{1/4}} \, e^{4\beta R_{\star} / v_w-I_{\star} \, e^{\beta R_{\star}/v_w}} \left( 1 - e^{-I_{\star}e^{\beta R_{\star}/v_w}} \right)\\
    &&\quad \times \Theta\left(M_{\rm PBH}-7.2\times10^{-3} \, \sqrt{g_{\phi} + \frac{7}{8}g_{\chi}} \, \frac{\alpha^2_\Lambda}{v_w} \, \frac{M^4_{\rm Pl}}{T_\star\Lambda^2}\right),
\label{eq:PBHDis2}
\end{eqnarray}
where $\Theta(.)$ is the Heaviside step function.

\begin{figure}[t]
  \centering
\includegraphics[width=0.6 \textwidth]{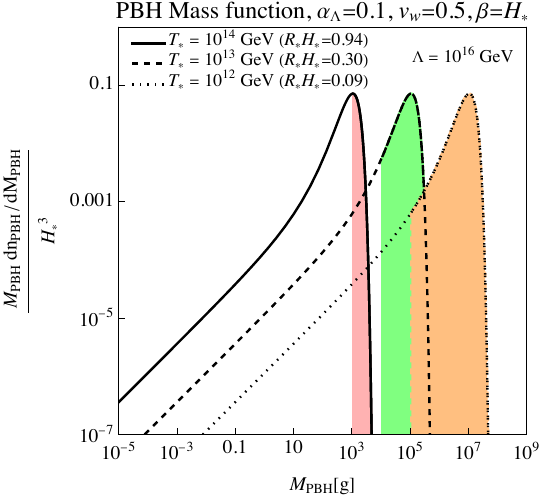}
  \caption{The PBH mass function for three different phase transition temperatures: $T_\star=10^{14}\,{\rm GeV}, T_\star=10^{13}\,{\rm GeV}, {\rm and} \, 
 T_\star=10^{12}\,{\rm GeV}$ as black solid, black dashed, and black dotted curves. The regions above the lower bound of PBH mass in Eq.~(\ref {eq:PBHDis2}) are shaded in red, green, and orange for the three benchmarks. The minimum $R_\star$ value for each benchmark is shown in the legend. The other parameters are set to be $\alpha_\Lambda=0.1$, $v_w=0.5$, $\beta=H_\star$, and $\Lambda =10^{16}\,{\rm GeV}$.}
\label{fig:massfun1}
\end{figure}

Fig.~\ref{fig:massfun1} shows the PBH mass function for three different phase transition temperatures: $T_\star=10^{14}\,{\rm GeV}, T_\star=10^{13}\,{\rm GeV}, T_\star=10^{12}\,{\rm GeV}$ as black solid, black dashed, and black dotted curves, and 
the lower bounds on the PBH mass are shaded in red, green, and orange respectively.
The other parameters are set to be $\alpha_\Lambda=0.1$, $v_w=0.5$, $\beta=H_\star$, and $\Lambda =10^{16}\,{\rm GeV}$. In the legends, we show the minimal values of $R_\star$ below which PBH formation is halted. As one expects from Eq.\eqref{eq:PBHDis2}, the PBH mass distribution is more peaked with a larger $R_\star H_\star$ value according to Eq.\eqref{eq:RstarMin}, as is shown with the red shaded distribution for $T_\star=10^{14}~{\rm GeV}$ and $\Lambda=10^{16}~{\rm GeV}$.

\section{UHDM from Black Hole Evaporation} \label{uhdmfrompbh}

In the previous Sections, we have discussed the formation of the PBHs and their associated mass function. In this Section, we discuss their evaporation and calculate the relic density of the produced UHDM candidates $\chi$ and $\phi$.

\subsection{Hawking Evaporation of PBHs}
PBHs, similar to all other black holes, lose their masses continuously via Hawking evaporation, regardless of formation mechanism~\cite{Hawking:1975vcx}. Hawking radiation emitted by black holes includes all the particles in the spectrum which are lighter than the instantaneous horizon temperature of the black hole. This temperature is inversely proportional to PBH mass:
\begin{equation}
T_\text{PBH}(t)=\frac{M_\text{Pl}^2}{8\pi M_\text{PBH}(t)}.
\label{eq:temp}
\end{equation}
While one can assign a temperature to a PBH, Hawking radiation is not described fully by a black body spectrum. The deviation of the Hawking radiation from the black body spectrum is coded in graybody factors~\cite{Page:1976df}. After ignoring these factors, the energy spectrum of the $i$th emitted species
of mass $m_i$ with $g_i$ degrees of freedom by a Schwarzschild black hole is given by:
\begin{equation}
\frac{d^2u_i(E,t)}{dtdE}=\frac{g_i}{8\pi^2}\frac{E^3}{e^{E/T_\text{PBH}(t)}\pm1},
\label{eq:rate}
\end{equation} 
($+/-$ for fermion/boson emission) where $E$ is the energy of the emitted particle, and $u_i(E,t)$ is the radiated energy per unit area of the black hole.

The evolution of a PBH mass with the initial value of $M_\text{PBH}(t_\star)$ at the formation time $t=t_\star$ is given by:
\begin{equation}
M_\text{PBH}(t)=M_\text{PBH}(t_\star)\left(1-\frac{t-t_\star}{\tau_\text{PBH}}\right)^{1/3},
\end{equation}
where
\begin{equation}
\tau_\text{PBH}=\frac{10240\pi}{g_\star(T_\text{PBH})}\frac{M_\text{PBH}^3(t_\star)}{M_\text{Pl}^4},
\end{equation}
denotes the lifetime of the PBH, and $g_\star(T_\text{PBH})$ is the number of relativistic degrees of freedom at temperature $T_\text{PBH}$ for the Hawking radiation process.

The rate of emission of the $i$th particle per energy interval is related to Hawking spectrum as:
\begin{equation}
\frac{d^2N_i}{dtdE}=\frac{4\pi r_\text{S}^2}{E}\frac{d^2u_i}{dtdE},
\label{eq:numrate}
\end{equation}
where $ r_\text{S}=2M_\text{PBH}/M_\text{Pl}^2$ is the 
Schwarzschild radius of the PBH.
The total number of bosonic ($B$) $i$th particles, emitted by one PBH  is obtained as:
\begin{eqnarray}
N_i&=&\frac{120\,\zeta(3)}{\pi^3}\frac{g_i}{g_\star(T_\text{PBH})}\frac{M_\text{PBH}^2(t_\star)}{M_\text{Pl}^2},~~~~~~T_\text{PBH}(t_\star)>m_i,
\label{eq:numberlight}\\
N_i&=&\frac{15\,\zeta(3)}{8\pi^5}\frac{g_i}{g_\star(T_\text{PBH})}\frac{M_\text{Pl}^2}{m_i^2},~~~~~~~~~~~~~~~T_\text{PBH}(t_\star)<m_i.
\label{eq:numberheavy}
\end{eqnarray}
The total number of fermionic species ($F$) is $N_F=\frac{3}{4}\frac{g_F}{g_B}N_B$.

\subsection{PBH Evaporation in a Radiation-Dominated Era}
DM, as a particle in the spectrum can also be emitted by Hawking evaporation of PBHs.
If PBHs never dominate the energy density of the Universe, then the injected entropy in the Universe at evaporation time is negligible. Assuming the emitted DM particles by PBHs do not annihilate each other after emission (See subsection~\ref{subsec:annihiltion} for validity of this assumption),  the yield of DM today, at $t_0$, is related to the yield of DM at evaporation time, $t_\text{eva}$:
\begin{equation}
    Y_{\rm DM}=\frac{n_{\rm DM}(t_0)}{s(t_0)}=\frac{n_{\rm DM}(t_\text{eva})}{s(t_\text{eva})}\simeq N_{\rm DM}\frac{n_\text{PBH}(t_\star)}{s(t_\star)},
    \label{eq:YRD1}
\end{equation}
where $n_{\rm DM}(t)$ and $n_\text{PBH}(t)$ are the number densities of DM particles and PBHs, respectively. $N_{\rm DM}$ is the total number of DM particles emitted by one PBH given by Eqs.~(\ref{eq:numberlight}) and (\ref{eq:numberheavy}). 
$s(t)$ is the entropy density evaluated as:
\begin{equation}
s(T)=\frac{2\pi^2}{45}g_{*,s}(T)T^3,~~~~~~g_{*,s}(T)=\sum_B g_B\left(\frac{T_B}{T}\right)^3+\frac{7}{8}\sum_F g_F\left(\frac{T_F}{T}\right)^3.
\label{eq:entropy}
\end{equation}
The yield of DM is proportional to the number density of PBHs, and consequently is proportional to the fraction of energy density of the Universe which collapsed into PBHs at the formation time, $\beta_{\rm PBH}=d\beta_{\rm PBH}(M_{\rm PBH})/d\ln M_{\rm PBH}$. DM yield in terms of initial abundance of PBHs is given by: 
\begin{equation}
    Y_{\rm DM}=\beta_{\rm PBH} N_{\rm DM}\frac{1}{M_{\rm PBH}}\frac{\rho_\text{rad}(t_\star)}{s(t_\star)}\simeq \frac{3}{4}\beta_{\rm PBH} N_{\rm DM} \frac{T_{\star}(M_{\rm PBH})}{M_{\rm PBH}},
    \label{eq:YRD2}
\end{equation}
where $\rho_\text{rad}(t_\star)$ and $T_\star(M_{\rm PBH})$ are the energy density and the temperature of the Universe at the PBH formation time respectively, and we assumed $g_\star(T_\star)\simeq g_{\star,s}(T_\star)$. The relationship between PBHs' mass and the temperature of the Universe at the formation time is dictated by PBH formation mechanism. 

Finally, the relic abundance of DM is obtained as:
\begin{equation}
\Omega_{\rm DM}=\frac{\rho_{{\rm DM}(t_0)}}{\rho_c}=\frac{m_{\rm DM} Y_{\rm DM} }{\rho_c}s(t_0),
\label{eq:relicdef}
\end{equation}
where $\rho_c(t_0)=1.0537\times 10^{-5}\, h^2\,\,\rm{GeV}~{\rm cm}^{-3}$, $s(t_0)=2891.2\left({T_0}/{2.7255 {\rm K}}\right)^3 \rm{cm^{-3}}$, and $h=0.674$ is scaling factor for Hubble expansion rate~\cite{Planck:2018vyg}.

Since $\rho_{\rm PBH}(t)/\rho_{\rm rad}(t)\propto a$, PBHs formed in a radiation-dominated era may lead to an early matter-dominated epoch before their evaporation. The critical initial abundance of PBHs, $\beta_c$, that can lead to an early matter-dominated epoch is related to the temperature of the Universe at the formation time, $T_\star$ through:
\begin{equation}
\frac{\rho_\text{PBH}(T_\text{early-eq})}{\rho_\text{rad}(T_\text{early-eq})}=\beta_c\frac{T_\star}{T_\text{early-eq}}\sim 1.
\end{equation}
An early matter-dominated epoch caused by PBHs is equivalent to $\beta_{\rm PBH}\gtrsim \beta_c$ where $\beta_c=T_{\rm eva}/T_\star$ and $T_{\rm eva}$ is the temperature of the Universe at PBH evaporation time.

\subsection{UHDM from PBHs Formed by a Trapped Dark Sector}
The mass of the PBHs formed by a trapped dark sector when the temperature of the Universe is $T_\star$, i.e., the temperature of the phase transition, is given by Eq.~\eqref{eq:MPBH} as:
\begin{equation}
M_{ \rm PBH} = \left[\frac{2 \pi^3}{90} \left(g_{\phi} + \frac{7}{8}g_{\chi} \right)\right]^{1/2} R_{\star}^{2} \, T_{\star}^{2} \, M_{\rm Pl}.
\label{eq:PBHmass1}
\end{equation}
It is easy to see that since we are interested in $m_\phi=m_\chi> T_\star$, the dark sector particles are always heavier than the initial temperature of the PBHs. 

We are now in a position to compute the relic density of DM today. We note that the portal operator in $\mathcal{L}_{\rm SM-DS}$ can induce the decay of $\phi$ to SM particles in true vacuum regions, while $\chi$ remains as a stable DM candidate. Indeed, the portal operator in Eq.\eqref{eq:EFTLambda} induces the decay of $\phi$ into a pair of Higgs bosons through the $\chi$-loop in true vacuum, and the decay width is $\Gamma_{\phi\to hh}\sim  y^2_\chi \, \alpha^2_\Lambda \, m^3_\phi / (16 \, \pi^2 \, \Lambda^2)$. Unlike in the true vacuum, thermal masses prevent $\phi$ decay in the false vacuum during PBH formation. In this case, $\phi$ will decay soon after being emitted by the PBH, and $\chi$ will be the only DM component in our benchmark dark sector model. However, for the relic density computation, we will present general results for the case where  both $\chi$ and $\phi$ are stable DM candidates, in anticipation of models where the portal operator is chosen such that $\phi$ is stable.   We will also discuss the limit that one should take when $\phi$ decays early.

The contribution of dark sector particles emitted by PBHs into the DM abundance today is obtained as:
\begin{eqnarray}
\nonumber \Omega_{\rm DM}&=&\Omega_{\chi}+\Omega_{\phi}\\
&=&\frac{45\sqrt{3}\times5^{1/4}\zeta(3)}{16 \times 2^{1/4}\pi^{23/4}}\frac{(3g_{\chi}+4g_{\phi})(7g_{\chi}+8g_{\phi})^{1/4}}{(8g_\star+7g_{\chi}+8g_{\phi})^{3/2}} \beta_{\rm PBH} \frac{R_\star}{H_\star^{-1}}\frac{s(t_0)}{m \rho_c}\frac{M_{\rm Pl}^{7/2}}{M_{\rm PBH}^{3/2}},
\label{eq:OmegaDM}
\end{eqnarray}
where $m_\chi=m_\phi=m$. We assumed the PBH mass function is monochromatic and take the $R_\star$ value from the minimal pocket radius in Eq.~\eqref{eq:RstarMin}. In the case when the scalar particle $\phi$ decays before BBN, the Eq.\eqref{eq:OmegaDM} can be modified by replacing the $3 g_\chi+4 g_\phi$ term in the numerator with $3 g_\chi$ such that only the relic abundance of stable $\chi$ particles contributes to DM. The remaining $g_\phi$ dependence arises from the existence of $\phi$ throughout the PBH formation and evaporation, which is irrelevant to its final decay.

\begin{figure}[t]
  \centering
\includegraphics[width=0.5 \textwidth]{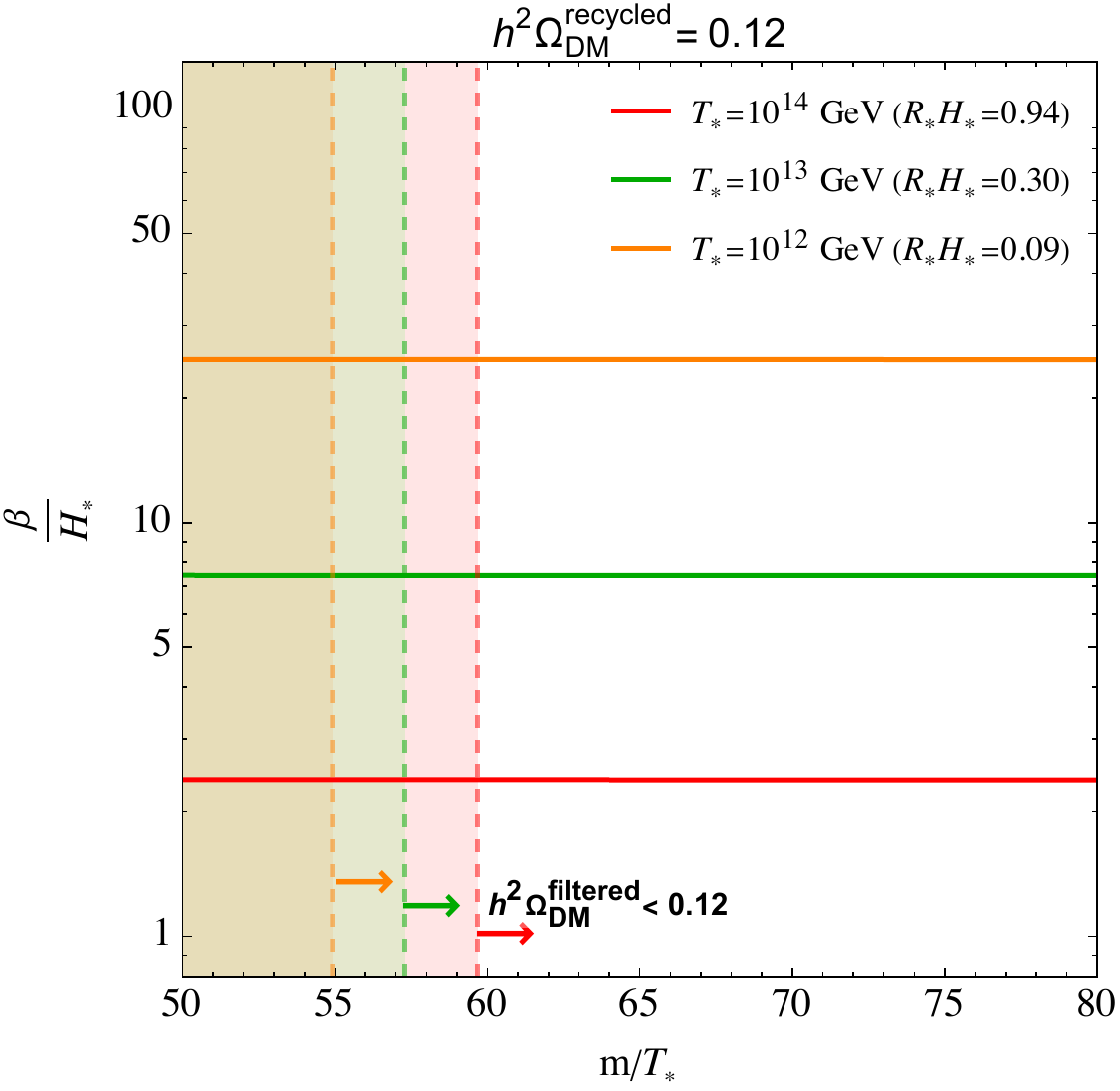}
  \caption{Value of the FOPT parameter $\beta$ that generates the observed DM relic abundance is shown with solid curves. The curve colors correspond to the benchmarks used in Fig.\ref{fig:massfun1} for different FOPT temperatures. The vertical dashed curves delineate regions where the filtered DM abundance is smaller than the observed abundance. Regions to the left of the dashed curves are excluded.}
\label{fig:betaFOPTmOverT}
\end{figure}

In Fig. \ref{fig:betaFOPTmOverT}, we show the FOPT parameters that allows the correct PBH formation rate $\beta_{\rm PBH}$ for different DM masses $m_\chi$, which later produce the observed DM relic abundance $\Omega_{\rm DM} h^2=0.12$ through recycling. The FOPT temperatures are chosen $T_\star = 10^{14}~{\rm GeV}$ (red), $10^{13}~{\rm GeV}$ (green), and $10^{12}~{\rm GeV}$ (orange) corresponding to the benchmark values used in Fig.\ref{fig:massfun1} with the same filling colors. The bubble wall velocity is $v_w=0.5$, and the $R_\star$ values are determined with $\alpha_\Lambda=0.1$ and $\Lambda=10^{16}~{\rm GeV}$. The PBH mass is determined with $R_\star$ assuming the mass function is monochromatic, and we approximate the $\beta_{\rm PBH}$ with its peak value. The $\beta/H_\star$ is almost a constant because of the exponential dependence in Eq.~\eqref{eq:PBHDis2}. Slow FOPTs with smaller $\beta$ values are required for recycled DM to be produced by small PBHs formed during high temperature phase transitions. We also show the region of $m/T_\star$ where the filtered contribution $\Omega_{\rm DM}^{\rm filtered}$, consisting of $\chi$ and $\phi$ particles that penetrated into the true vacuum during the FOPT, is a subdominant fraction of the total DM abundance. The filtered DM abundance soon becomes negligible to the right of the dashed curves since the filtering rate is exponentially suppressed.
 
\begin{figure}[h]
  \centering
\includegraphics[width=0.48\textwidth]{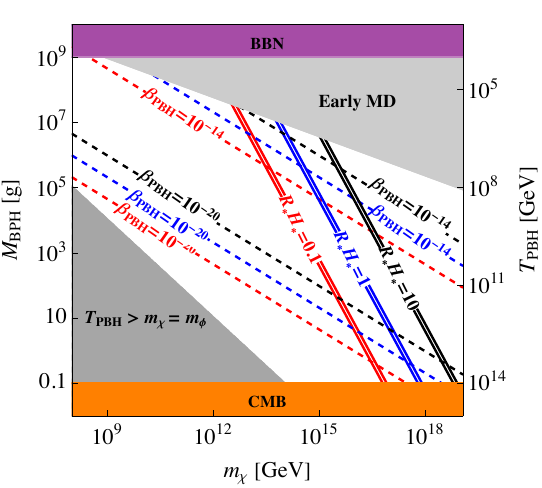}\quad
\includegraphics[width=0.48\textwidth]{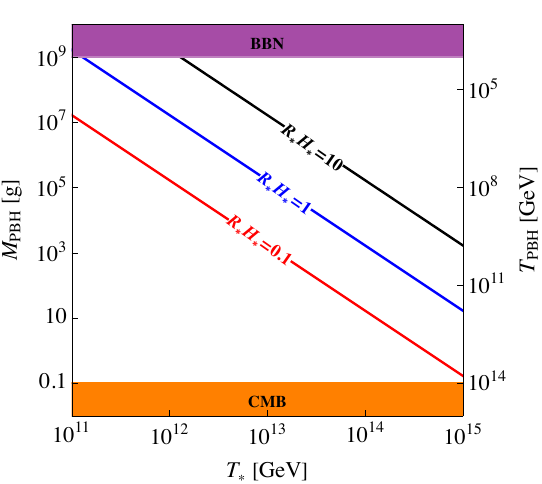}
  \caption{Constraints on dark sector particles produced by PBH evaporation. \textbf{Left:} Dashed
lines show the upper limits on $\beta_{\rm PBH}$ assuming PBHs evaporation explains all of the DM today. The orange and purple regions are excluded by constraints from the CMB and BBN, respectively. The dark gray region correspond to PBHs with an initial temperature larger than the mass of the DM particles. Within the light gray area, 
$\beta_{\rm PBH}$ is so large that it leads to an early matter-dominated era, and the abundance of
DM particles produced by PBH evaporation is independent of $\beta_{\rm PBH}$.
Different colors of dashed contours correspond to different choices of $R_{\star}H_{\star}$: red, blue, and black for $0.1$, $1$, and $10$ respectively.
Solid lines with the same color show the region of the parameter space within which $50\lesssim m/T_\star\lesssim 65$ and consequently the filtered component is suppressed. The left side of the red solid lines corresponds to $R_{\star}H_{\star}<0.1$ which gives rise to an extended mass function for PBHs and is not compatible with the calculations in this study which are based on a monochromatic mass function for PBHs.
On the right side of the black solid lines, the mass of the dark sector particles become closer to the Planck scale, and  the validity of the portal model from the point of view of effective field theory becomes less reliable. \textbf{Right:} PBH mass as a function of $T_\star$ for different choices of $R_{\star}H_{\star}$.}
  \label{fig:parameterspace}
\end{figure}

Our primary results are shown in Fig.~\ref{fig:parameterspace}. The left panel depicts the upper limits on the initial abundance of PBHs, $\beta_{\rm PBH}$, on the $(M_{\rm PBH},m_\chi)$ parameter space assuming $\Omega_\chi=\Omega_{\rm DM}$, as well as excluded regions. The excluded PBH masses by BBN and CMB are shaded in purple and orange, respectively. Within the dark gray area, the initial temperature of PBHs is larger than the mass of the DM particles. 
The dashed contours represent the upper limits on $\beta_{\rm PBH}$ when PBHs produce all the
DM after evaporation in a radiation-dominated universe.
In the light gray area the required initial abundance of PBHs to produce the observed abundance of DM today, is larger than $\beta_c$ and therefore gives rise to an early matter-dominated era.  
With the monochromatic assumption, the PBH mass is determined by the minimal pocket radius $R_{\star}$ and the FOPT temperature $T_\star$. Different colors of dashed contours correspond to different choices of $R_{\star}H_{\star}$: red, blue, and black for $R_{\star}H_{\star}=0.1$, $R_{\star}H_{\star}=1$, and $R_{\star}H_{\star}=10$ respectively.
Solid lines with the same color show the region of the parameter space within which $50\lesssim m/T_\star\lesssim 65$ and consequently the filtered component is suppressed. The left side of the red solid lines corresponds to $R_{\star}H_{\star}<0.1$ which leads to an extended mass function for PBHs and is not compatible with the calculations in this study which are based on a monochromatic mass function for PBHs.
On the right side of the black solid lines, the mass of the dark sector particles become closer to the Planck scale, and since there is another scale between the vev of the dark scalar and the Planck scale, i.e., $\Lambda$, the validity of the portal model becomes less reliable. The right panel of Fig.~\ref{fig:parameterspace} shows PBH mass as a function of $T_\star$ for different choices of $R_{\star}H_{\star}$. 

\subsection{Annihilation of Dark Matter after Evaporation of PBHs}
\label{subsec:annihiltion}
If the number of DM particles emitted by PBHs are large enough, they may annihilate each other and therefore have a smaller final abundance. To examine this possibility, we need to evaluate the rate of annihilation of DM particles and compare that with the Hubble expansion rate at the evaporation time (and maybe afterwards, when produced particles are ultra relativistic).

The ratio of annihilation rate of DM particles emitted by PBHs to the Hubble expansion rate, immediately after PBH evaporation, at $t=\tau_{\rm PBH}+\epsilon$, can be estimated as:
\begin{eqnarray}
    \frac{\Gamma_{\rm ann}(\tau_{\rm PBH}+\epsilon)}{H(\tau_{\rm PBH}+\epsilon)}&\simeq &
    \frac{n_\chi(\tau_{\rm PBH}+\epsilon)\langle\sigma v\rangle_{\chi\chi}(\tau_{\rm PBH}+\epsilon)}{H(\tau_{\rm PBH}+\epsilon)},
\end{eqnarray}
where
\begin{eqnarray} 
    \nonumber n_\chi(\tau_{\rm PBH}+\epsilon)=N_\chi n_{\rm PBH}(\tau_{\rm PBH}-\epsilon)&=&N_\chi n_{\rm PBH}(t_\star)\frac{a^3(t_\star)}{a^3(\tau_{\rm PBH})}\\
   \nonumber &=&N_\chi \beta_{\rm PBH} \frac{\rho_{\rm rad}(t_\star)}{M_{\rm PBH}}\frac{a^3(t_\star)}{a^3(\tau_{\rm PBH})}\\
   &\simeq&N_\chi \beta_{\rm PBH} \frac{\rho_{\rm rad}(t_\star)}{M_{\rm PBH}}\frac{T^3(\tau_{\rm PBH})}{T^3(t_\star)},
\end{eqnarray}
$T(\tau_{\rm PBH})$ is the temperature of the Universe at evaporation time, and annihilation cross-section is given by:
\begin{equation}
 \langle\sigma v\rangle_{\chi\chi}(\tau_{\rm PBH}+\epsilon)\sim \frac{\alpha_\Lambda^2}{{\rm max} [\bar E_\chi^2, \Lambda^2]}=\frac{\alpha_\Lambda^2}{\Lambda^2},
\end{equation}
where $\bar E_\chi$ is the average energy of $\chi$ particles emitted by PBHs which for $m_\chi>T_\text{PBH}$ is given by $\bar E\simeq 5.4 m_\chi$~\cite{Gondolo:2020uqv,Sandick:2021gew}.

By using the lifetime of PBHs and Friedmann equation, one can find the temperature of the Universe at evaporation time to be equal to:
\begin{equation}
T(\tau_{\rm PBH})=\,\frac{\sqrt{3}\,g^{1/4}_\star(T_\text{PBH})}{64\sqrt{2}\,5^{1/4}\pi^{5/4}}\frac{M_\text{Pl}^{5/2}}{M_\text{PBH}^{3/2}}.
\label{eq:Trevap}
\end{equation}

Therefore, we obtain
\begin{eqnarray}
      \nonumber  \frac{\Gamma_{\rm ann}(\tau_{\rm PBH}+\epsilon)}{H(\tau_{\rm PBH}+\epsilon)}&\simeq&
        \frac{27\sqrt{2} \zeta(3)}{33554432\times 2^{1/4}\sqrt{5}\pi^{15/2}}
        \frac{\sqrt{g_\star}}{(7g_\chi+8g_\phi)^{1/4}}\frac{\alpha_\Lambda^2 \beta_{\rm PBH}}{R_\star H_\star}\frac{M_{\rm Pl}^9}{M_{\rm PBH}^5 m_\chi^2\Lambda^2}\\
       \nonumber &<&
        \frac{27\sqrt{2} \zeta(3)}{33554432\times 2^{1/4}\sqrt{5}\pi^{15/2}}
        \frac{\sqrt{g_\star}}{(7g_\chi+8g_\phi)^{1/4}}\frac{\alpha_\Lambda^2 \beta_{\rm PBH}}{R_\star H_\star}\frac{M_{\rm Pl}^9}{M_{\rm PBH}^5 m_\chi^4}\\
        &<&
        \frac{27\sqrt{2} \zeta(3)}{8192\times 2^{1/4}\sqrt{5}\pi^{7/2}}
        \frac{\sqrt{g_\star}}{(7g_\chi+8g_\phi)^{1/4}}\frac{\alpha_\Lambda^2 \beta_{\rm PBH}}{R_\star H_\star}\frac{}{}\frac{M_{\rm Pl}}{M_{\rm PBH}}\ll 1,
\end{eqnarray}
where in the second step we used $\Lambda > m_\chi$, and in the third step we applied the condition that $T_{\rm PBH}=M_{\rm Pl}^2/(8\pi M_{\rm PBH})< m_\chi$. 

Since $n_\chi\propto a^{-3}$, and $H\propto a^{-2}$, $n_\chi\langle\sigma v\rangle_{\chi\chi}/H$ attains its  maximum  value immediately after evaporation and then decreases by the expansion of the Universe.
It is clear that the rate of annihilation is always less than the Hubble rate and therefore, the yield of DM particles produced by PBHs does not change after production by PBHs. Although the DM candidates produced by PBHs do not establish chemical equilibrium with the SM bath, but they may reach kinetic equilibrium with it again due to elastic scattering processes.

\section{Discussions and Conclusions}

In this work, we studied a new mechanism for producing UHDM: recycling. While we reserve a study of the phenomenology of recycled DM for the future, there are some general novel features that we comment on. 

In recycling, multiple DM candidates reside, as massless particles, in a dark sector which is initially in kinetic equilibrium with the SM. The dark sector undergoes a FOPT, and the DM candidates acquire a mass much larger than their temperature and therefore get trapped in the false vacuum. The rate of leakage of dark sector particles into the true vacuum is suppressed enough that eventually (some of or even all of) the pockets of false vacuum collapse into PBHs. If the FOPT occurs at high temperatures, the formed PBHs would be light enough to evaporate before BBN and escape all the cosmological constraints. During Hawking evaporation, PBHs reproduce (recycle) the DM candidates that initially collapsed into them. The DM particles emitted by PBHs are ultra-heavy, due to being materialized in the true vacuum, and are sparse enough to be able to explain the observed abundance of DM today without overclosing it. The suppression of DM annihilation into SM dictated by very heavy mediators leads to a lower bound on the size of the false vacuum pockets that can collapse into PBHs successfully. Therefore, the PBH mass function, in general, is an extended distribution, with a low mass cutoff determined by the IR physics. In the case that most (or all) of the pockets collapse into PBHs, we expect an early PBH dominated Universe, since almost all the energy content of the dark sector which is of the order of percentage of the radiation energy density ends up in PBHs. It is worth mentioning that the high temperature of FOPT connected to recycling DM by PBHs gives rise to the gravitational wave signatures in the ultra high frequency MHz-GHz range~\cite{Gehrman:2023esa, Gehrman:2022imk}. 

In this study, we focused on scenarios within which the lower bound on the mass of PBHs is not too small and therefore the resultant PBH mass function can be treated as a monochromatic distribution. Generally speaking, during the FOPT, a portion of the dark sector energy may leak into the true vacuum, while the rest gets trapped in the false vacuum. When the mass of the DM in true vacuum is larger than the phase transition temperature by a factor of $\sim$ 55 or more, then the filtered DM contribution into the final abundance of DM becomes negligible and the recycled component emitted by PBHs afterwards would be the dominant constituent. Recycling, via formation of PBHs and their subsequent Hawking evaporation preceded by a FOPT, provides a plausible way to produce UHDM which has been in thermal equilibrium with the SM in early Universe. Depending on details of the FOPT and the particle physics model, UHDM with a mass heavier than $\sim 10^{12}\,{\rm GeV}$ and the right relic abundance today, can be produced by recycling mechanism initiated with FOPTs occurring at temperatures larger than $\sim 10^{11}\,{\rm GeV}$. Since recycling mechanism involves a dark sector which interacts with the SM, a FOPT succeeded by emission of high frequency gravitational waves, and formation of PBHs with interesting mass functions, it provides a novel UHDM production mechanism with a rich phenomenology which requires further study in the future. 

We comment on future improvements of the analysis. The bubble wall motion that drives the overdensity necessary for PBH formation is influenced by the dynamics of the dark sector and its back-reaction on the wall, as recently studied in \cite{Lewicki:2023mik} with simulations for non-interacting dark sector species in the false vacuum pocket. The recycled DM model consists of a thermalized dark sector with substantial self-interactions through the Yukawa force mediated by light scalar $\phi$. In this study, we assume constant wall velocities to showcase the recycled DM model, and leave a comprehensive exploration of bubble dynamics with the active Yukawa interaction to future works.

\vspace{11pt}
\noindent {\it Note added}: As we were finalizing this paper, we became aware of \cite{Kim:2023ixo}. While there is some overlap, our study emphasises UHDM recycling with PBHs formed from the direct collapse of the trapped dark sector, where its portal coupling to the SM controls the PBH mass distribution, as opposed to Fermi-ball formation.

\acknowledgments
The work of B.S.E. is supported in part by DOE grant DE-SC0022021. The work of K.S. and T.X. is supported in part by DOE grant DE-SC0009956.

\bibliographystyle{JHEP}
\bibliography{RDM}
\end{document}